\documentclass[aps,pra,showpacs,twocolumn,superscriptaddress,nolongbibliography,floatfix]{revtex4-2}
\usepackage{graphicx}
\usepackage{amsmath}
\usepackage{xparse}
\usepackage{dcolumn}
\usepackage{bm}
\usepackage[colorlinks=true, citecolor=blue,allcolors=blue]{hyperref}
\usepackage{physics} 
\usepackage{comment}
\usepackage{multirow}

\begin{document}
\title{A delay-programmable two-color femtosecond source for multiphoton ionization studies based on chirped-seed NOPA
}

\author{ K.~Foster}
\affiliation{Physics Department and LAMOR, Missouri University of Science \& Technology, Rolla, MO 65409, USA}
\author{ S.~Majumdar}
\affiliation{Physics Department and LAMOR, Missouri University of Science \& Technology, Rolla, MO 65409, USA}
\author{ M.~Toombs}
\affiliation{Physics Department and LAMOR, Missouri University of Science \& Technology, Rolla, MO 65409, USA}
\author{ H.~Agarwal}
\affiliation{Physics Department and LAMOR, Missouri University of Science \& Technology, Rolla, MO 65409, USA}
\author{ D.~Fischer}
\email[]{fischerda@mst.edu}
\affiliation{Physics Department and LAMOR, Missouri University of Science \& Technology, Rolla, MO 65409, USA}

\date{\today}
\begin{abstract}
We demonstrate a delay-programmable two-color femtosecond source based on a chirped-seed noncollinear optical parametric amplifier. Introducing controlled dispersion into the seed enables spectral selection through pump–seed delay, allowing flexible generation of two independently tunable pulse components with adjustable relative timing at high repetition rate. The temporal and spectral properties are characterized using nonlinear optical cross-correlation and dispersion-scan measurements. As a benchmark application, the source is employed in a COLTRIMS-based multiphoton ionization experiment on trapped Li atoms, revealing delay-dependent ionization pathways and demonstrating its suitability for bichromatic ultrafast spectroscopy.
\end{abstract}
\maketitle
\section{Introduction}

Multiphoton ionization driven by femtosecond laser pulses is a central tool for investigating atomic and molecular dynamics with high temporal resolution. In particular, bichromatic excitation schemes enable control over ionization pathways and access to interference between competing multiphoton channels in both atoms and molecules through quantum-path interference \cite{Shapiro2011, Eickhoff2021, Demekhin2019, Mezinska2024}. These experiments therefore require femtosecond light sources that provide independent control over spectral composition and relative timing of multiple pulse components, which remains challenging in conventional broadband amplification schemes.

A variety of approaches have been developed for generating tailored femtosecond waveforms with controlled spectral and temporal structure. Common strategies include Fourier-domain pulse shaping using spatial light modulators, nonlinear frequency conversion schemes such as second-harmonic or sum-frequency generation, and multi-channel architectures based on independent optical parametric amplifiers \cite{Weiner2000, Weiner2011, Boyd2020, Cerullo2003}. While these techniques provide significant flexibility, they are often limited in either pulse energy, spectral bandwidth, or the ability to independently and continuously tune multiple pulse components with precise temporal control. In particular, achieving stable two-color operation with simultaneous spectral programmability and adjustable inter-pulse delay remains experimentally challenging in high-repetition-rate systems.

To address these limitations, we implement a chirped-seed noncollinear optical parametric amplifier (NOPA) scheme \cite{harth2018compact}, enabling independent control over the spectral composition and relative timing of two amplified pulse components. The seed pulse is deliberately temporally stretched by introducing a controlled amount of dispersion prior to amplification, resulting in a strongly chirped pulse with a duration of approximately 1 ps. Due to the intrinsic time–frequency mapping of the chirped seed \cite{Dubietis1992, Cerullo2003}, different spectral components interact with the pump at different temporal delays within the NOPA gain window. Consequently, the central wavelength and bandwidth of the amplified output can be continuously tuned by adjusting the relative delay between pump and seed pulses. Here we demonstrate that this architecture simultaneously affords precise control over the inter-pulse delay between two independently tunable spectral components, making it well-suited for bichromatic strong-field experiments requiring both spectral selectivity and precise temporal control.

The underlying amplification concept employed in this work--chirped-seed noncollinear optical parametric amplification with delay-dependent spectral selection--has previously been used in our group as a stable femtosecond light source for photoelectron momentum spectroscopy experiments based on COLTRIMS \cite{Doerner2000, Ullrich2003, Fischer+2019+103+156}. In those studies, the laser system served primarily as a robust excitation source for investigations of circular and magnetic dichroism \cite{Silva2021,Silva2021b, Acharya2021, Acharya2022} as well as magnetic wavepacket dynamics \cite{Romans2025}, without a dedicated focus on the underlying laser architecture or its tunability. In contrast, the present work concentrates on the technical implementation and systematic characterization of the laser system itself, with particular emphasis on controlled two-color pulse generation and independent tuning of spectral and temporal degrees of freedom.


\section{Experimental Setup}

\begin{figure}[b]
    \centering
    \includegraphics[width=1\linewidth]{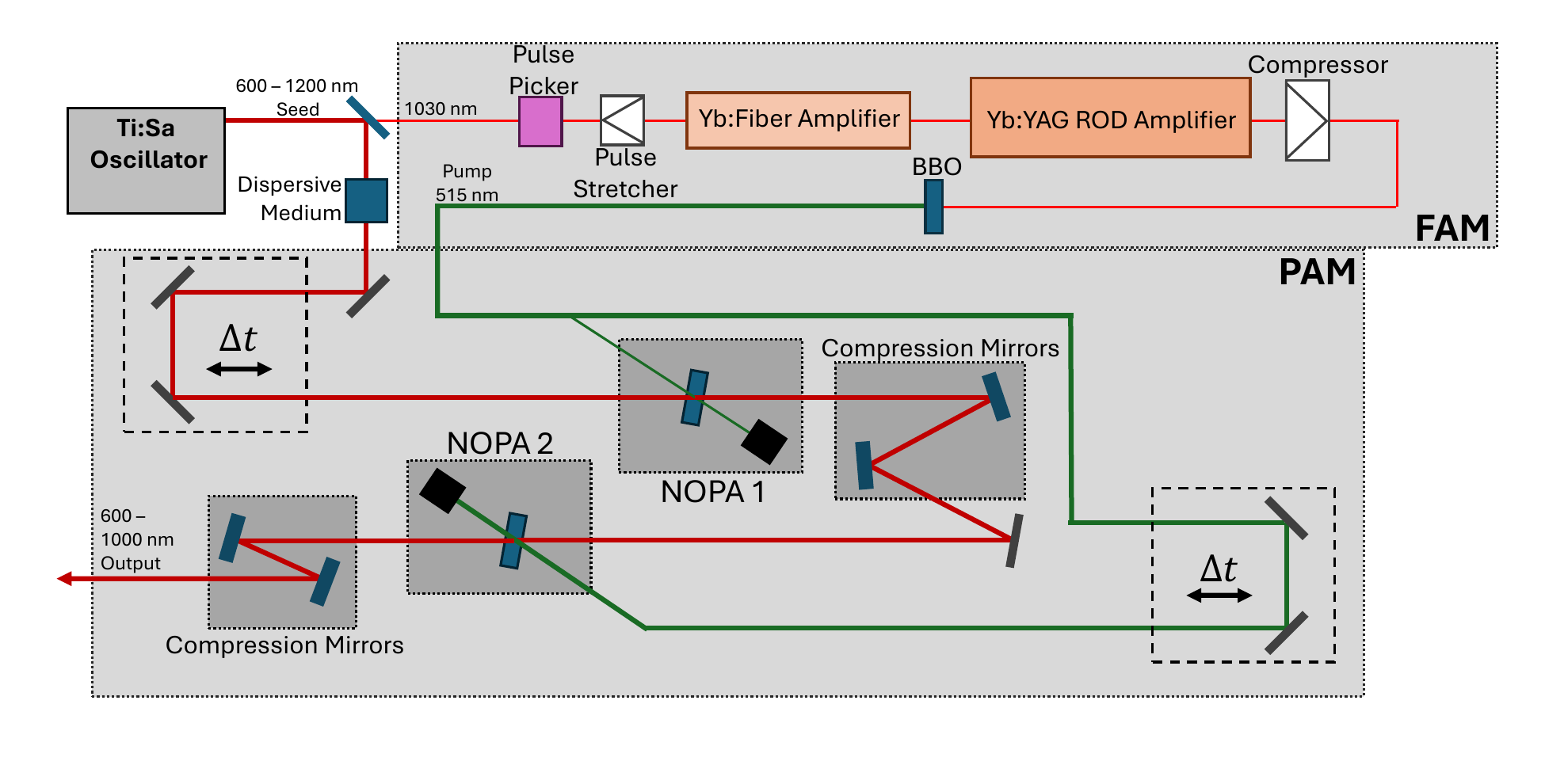}
    \caption{OPCPA laser system. In the original configuration of the laser no dispersive medium was present. In the modified configuration the dispersive medium was either 12.7\,mm of SF1 or a 5\,mm thick sapphire window at an incidence angle of $\approx$\,47.5$^\circ$ with respect to the surface normal.}
    \label{fig:fslaser}
\end{figure}

The laser system used in this study is based on a commercially available femtosecond optical parametric chirped-pulse amplifier (OPCPA) platform (Laser Quantum, similar to the model described in \cite{harth2018compact}). A schematic overview of the system is shown in Fig.~\ref{fig:fslaser}. In the following, we describe the main components of the system and highlight the modifications introduced for chirped-seed operation and two-color pulse generation.

\subsection{Original Configuration}

The system is seeded by a Ti:sapphire oscillator delivering broadband pulses spanning approximately 600–1200\,nm, with a pulse duration of $\sim$5\,fs, pulse energy of $\sim$2.5\,nJ, and a repetition rate of 80\,MHz.

A fraction of the oscillator output in the infrared (approximately 1020–1060\,nm) is used to generate the pump for the parametric amplification stages. This beam is amplified in a multi-stage fiber amplifier module (FAM), followed by repetition rate reduction to 200\,kHz using pulse pickers. Subsequent amplification in a rod-type amplifier and compression yields pulses of 150–200\,fs duration. These pulses are frequency-doubled in a second-harmonic generation (SHG) crystal to produce pump pulses at 515\,nm with energies of up to 90\,$\mu$J and an average power of approximately 18\,W.

The core of the laser system is the parametric amplifier module (PAM), in which the seed and pump pulses are spatially and temporally overlapped in two type-I BBO crystals of 2\,mm thickness. These stages operate in a noncollinear optical parametric amplification (NOPA) configuration \cite{Cerullo2003,Oien1997}, transferring energy from the pump to the seed while preserving the seed’s temporal structure. The seed beam passes sequentially through both stages (hereafter referred to as NOPA stage~1 and NOPA stage~2), while the pump beam is split by a thin-film polarizer. Approximately 20\,\% of the pump power is directed to the first stage and 80\,\% to the second. This ratio can be adjusted using a half-wave plate placed before the polarizer.

Both NOPA stages are arranged in a walk-off-compensated noncollinear geometry \cite{Oien1997}, with an angle of approximately $\theta = 24.4^\circ$ between the pump beam and the optical axis of the BBO crystal, and an angle of approximately $\alpha = 2.6^\circ$ between the seed and pump beams at the crystal. A small chirp is introduced in the seed beam after the first NOPA stage and is compensated by two reflections from a pair of double-chirped mirrors before entering the second stage. After the second stage, additional chirped mirror pairs and fused silica wedges are used to optimize pulse compression and achieve the shortest possible pulse duration.

It is worth noting that the seed beam, before the first NOPA stage, and the pump beam, for the second NOPA stage, pass through delay stages consisting of pairs of 45$^\circ$ mirrors mounted on motorized translation stages. These delay stages allow independent adjustment of the temporal overlap between the seed and pump pulses in the two NOPA stages. In the standard configuration, the system delivers an output beam with an average power of up to 3\,W, a central wavelength around 780\,nm, pulse durations as short as 8\,fs, a repetition rate of 200\,kHz, a pulse energy of up to 17\,$\mu$J, and a beam diameter of 1.2\,mm.

\subsection{Modified Operation}

In the present study, the commercial system was modified to enable selective amplification of narrow spectral regions of the broadband seed pulse. To this end, a dispersive element was introduced into the seed beam path immediately before the parametric amplifier module (PAM), as illustrated in Fig.~\ref{fig:fslaser}. This induces a significant chirp, stretching the seed pulse to a duration exceeding that of the pump pulse and thereby enabling delay-dependent spectral selection during parametric amplification \cite{Cerullo2003}.

In initial tests, a 5\,mm thick sapphire window was employed at an incidence angle of approximately 47.5$^\circ$ relative to the surface normal, close to the Brewster angle. This minimizes reflective losses while increasing the effective optical path length (to approximately 5.51\,mm). A drawback of this configuration is the introduction of angular dispersion (spatial chirp), resulting in a slight spatial variation of the spectral content across the beam profile.

\begin{figure}
    \centering
    \begin{tabular}{ |c||c|c|c|c|c| } 
        \hline
            & \shortstack{Ref.\\Index}
            & \shortstack{Glass\\Thickness\\(mm)} 
            & \shortstack{Pulse\\Delay\\(ps)} 
            & \shortstack{Pulse\\Broadening\\(fs)} 
            & \shortstack{GDD\\(fs$^2$)} \\
        \hline
         Sapphire & 1.76 & 5.56 & 460 & 259 & 320 \\
         PBS Cube & 1.70 & 12.7 & 424 & 1570 & 1930 \\  
        \hline
    \end{tabular}
    \caption{Calculated pulse parameters following propagation through either sapphire or a polarizing beam splitter (PBS) cube composed of N-SF1 glass, based on simulated material dispersion. The input pulse is assumed to have a duration of 4.4\,fs, a spectral bandwidth of approximately 600--1200\,nm.}
    \label{fig:table}
\end{figure}

\begin{figure}
    \centering
    \includegraphics[width=0.9\linewidth]{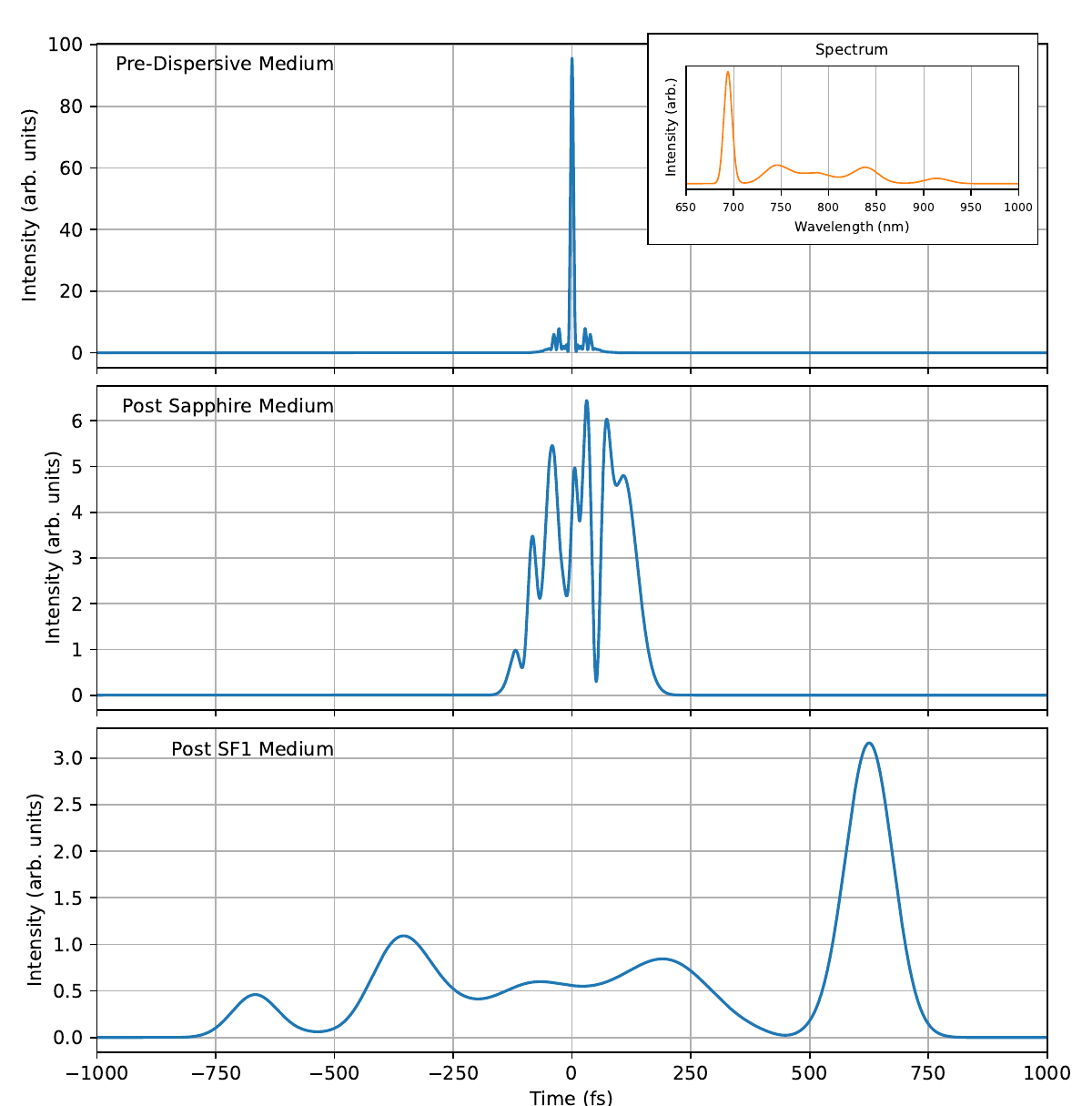}
    \caption{Simulated spectra and corresponding temporal pulses after propagation through dispersive media. (top) Initial 4.4\,fs input pulse and corresponding spectrum (inset). (middle) Propagation through 5\,mm of sapphire at an incidence angle of approximately 47.5$^\circ$. (bottom) Propagation through 12.7\,mm of N-SF1 glass in the PBS cube configuration.}
    \label{fig:pulsesim}
\end{figure}

In subsequent experiments, a polarizing beam splitter (PBS) cube (Thorlabs PBS122) was used as an alternative dispersive medium. The PBS, made of N-SF1 glass with a thickness of 12.7\,mm, introduces significantly larger group delay dispersion compared to the sapphire window, resulting in stronger temporal stretching of the seed pulse (see Fig.~\ref{fig:pulsesim}). This enhances the achievable spectral selectivity while maintaining compatibility with post-amplification pulse compression.

The selective spectral amplification is achieved by exploiting the time–frequency mapping of the chirped seed pulse in combination with adjustable pump–seed delay. By tuning the temporal overlap between pump and seed pulses within each NOPA stage, specific temporal slices of the chirped seed---and thus specific spectral components---are preferentially amplified. Independent delay stages in the seed path before the first NOPA stage and in the pump path before the second stage allow the two amplification stages to address either identical or distinct spectral regions of the seed pulse, enabling controlled generation of single- or two-color output pulses.

The parametric amplification process provides broadband gain within the phase-matching bandwidth of the NOPA stages. In the case of a strongly chirped seed pulse, this broadband gain is effectively temporally gated by the finite duration of the pump pulse. As a result, only a limited temporal---and therefore spectral---portion of the seed pulse is amplified for a given pump–seed delay. Increasing the seed pulse duration enhances this temporal separation of spectral components, allowing for narrower spectral selection.

%

\section{RESULTS}

 \begin{figure}
    \centering
    a)
    \includegraphics[width=0.9\linewidth]{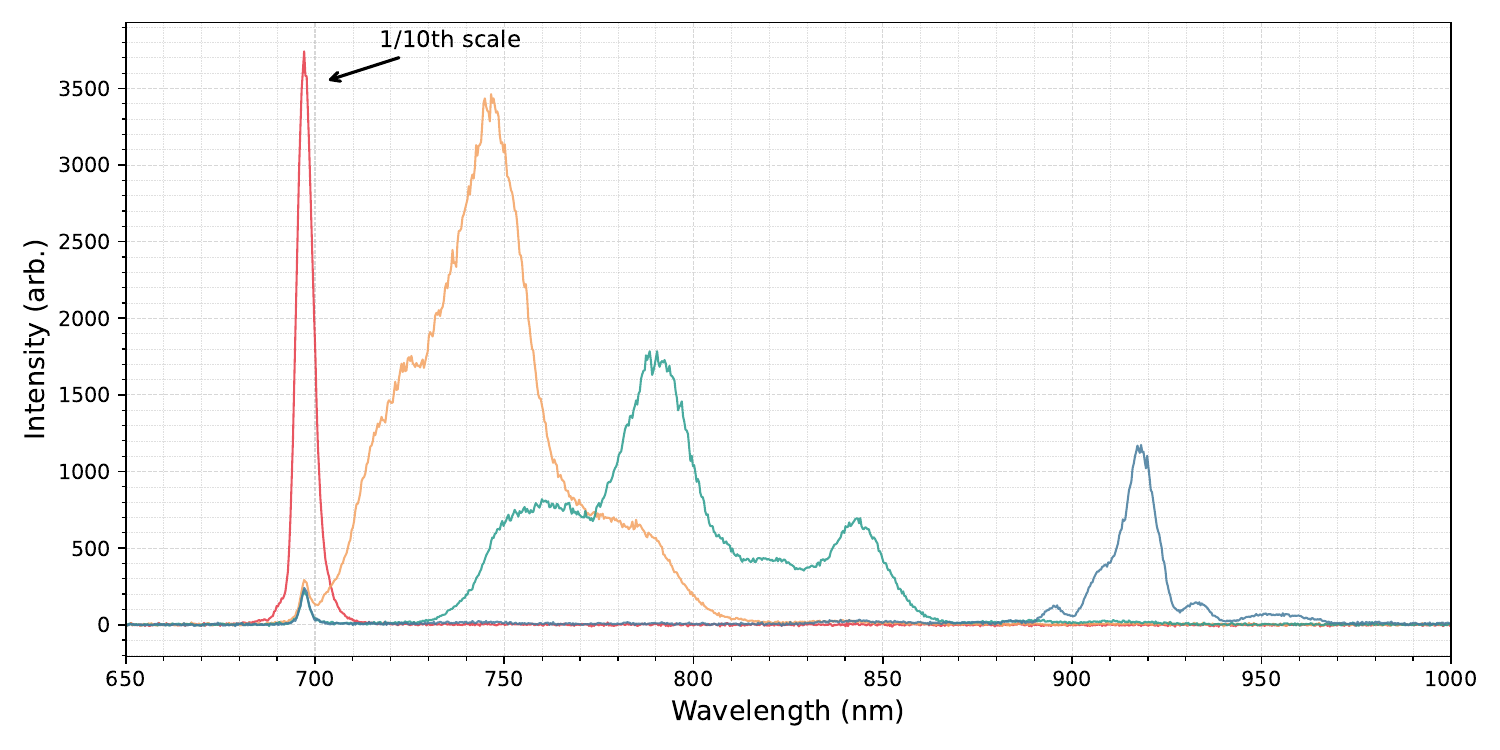}
    
    b)
    \includegraphics[width=0.9\linewidth]{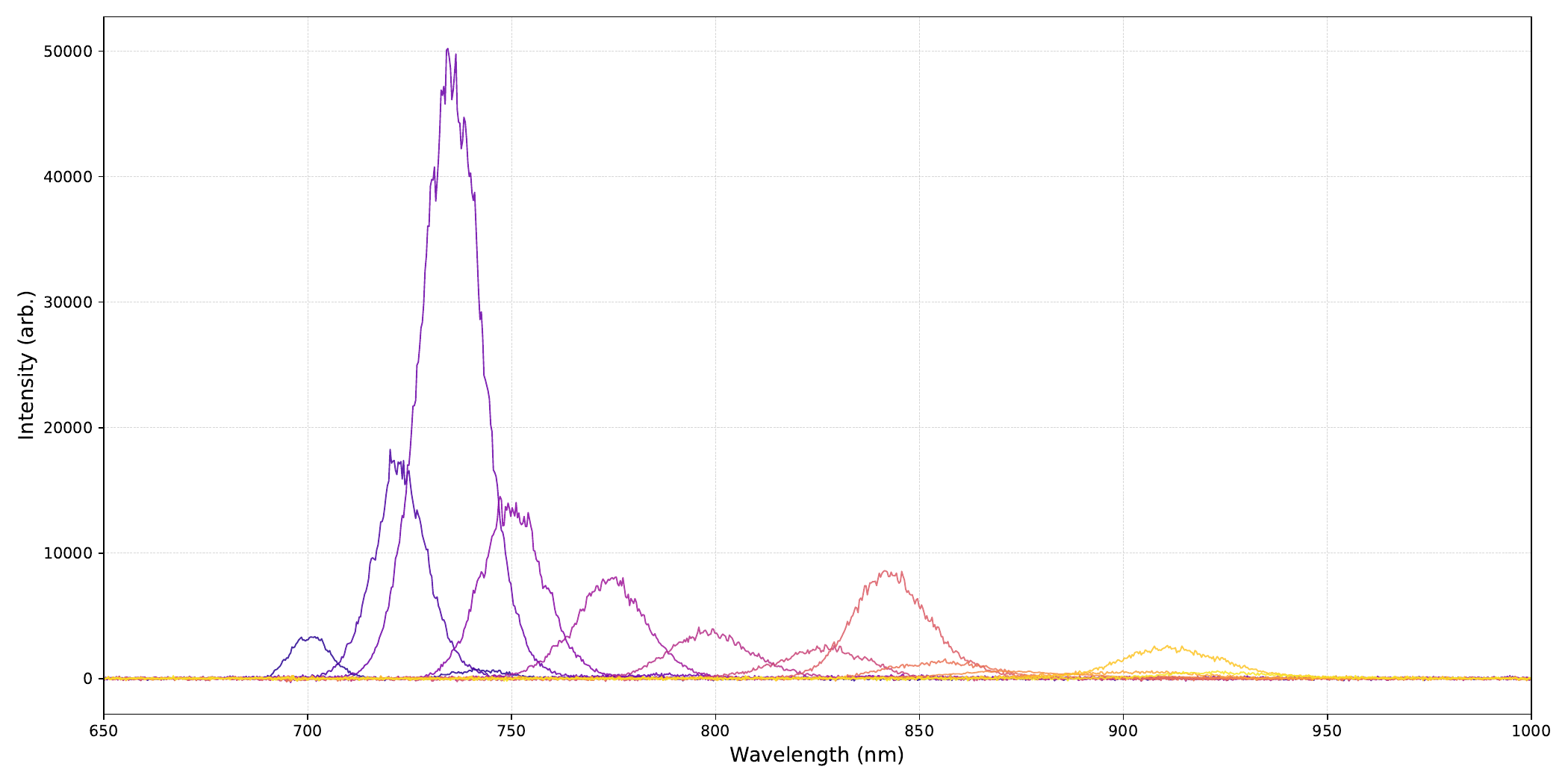}
    \caption{Output spectra of the OPCPA system obtained by tuning the delay stages of the two NOPA stages for different target wavelengths using (a) a 5.51\,mm sapphire window and (b) a 12.7\,mm N-SF1 PBS cube as dispersive media. For each scan, both delay stages were adjusted to maximize amplification at a selected spectral region, which was then varied between measurements. For wavelengths larger than 700\,nm, The PBS-based configuration yields significantly narrower and better-resolved spectral features compared to sapphire.}
    \label{fig:clrscans}
\end{figure}

The sapphire window introduces sufficient group delay dispersion (GDD) to enable frequency-selective amplification of the seed pulse, as shown in Fig.~\ref{fig:clrscans}\,(a). However, the resulting spectra remain relatively broad and only weakly structured. This is attributed to the limited temporal stretching of the seed pulse, which leads to incomplete time–frequency separation and thus reduced spectral selectivity during the parametric amplification process. As a result, only modest spectral discrimination is achieved, with partially resolved features near 690\,nm and 740\,nm.

In contrast, the PBS-based configuration produces clearly defined and significantly narrower spectral peaks, as shown in Fig.~\ref{fig:clrscans}\,(b). This improvement arises from the substantially larger group delay dispersion introduced by the SF1 glass, which results in stronger temporal stretching of the seed pulse and enhanced time–frequency separation.

\begin{figure}
    \centering
    \includegraphics[width=0.9\linewidth]{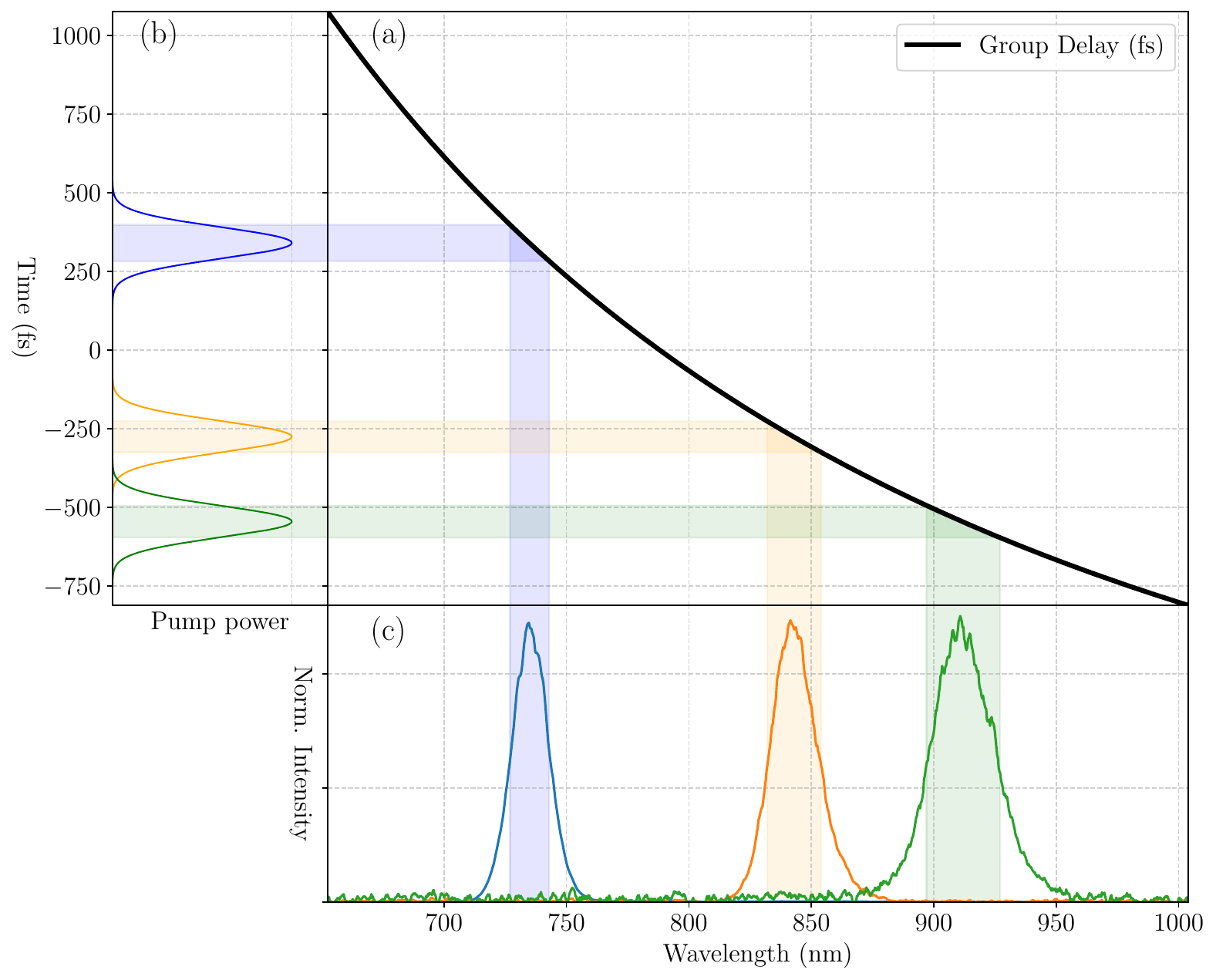}
    \caption{Illustration of delay-dependent spectral selection in the chirped-seed NOPA scheme. (a) Calculated group delay of a pulse after propagation through 12.7\,mm of SF1 glass. (b) Temporal position of the pump pulse for different delay settings. (c) Corresponding experimentally measured output spectra, showing the selection of different spectral components via pump–seed delay.}
    \label{fig:grpdelay}
\end{figure}

The underlying mechanism of this delay-dependent spectral selection is illustrated in Fig.~\ref{fig:grpdelay}. The group delay introduced by the SF1 medium maps the broadband seed spectrum onto the temporal domain (Fig.~\ref{fig:grpdelay}(a)). By adjusting the relative delay between pump and seed pulses (Fig.~\ref{fig:grpdelay}(b)), different temporal slices of the chirped seed are selectively amplified within the NOPA gain window. The corresponding measured spectra (Fig.~\ref{fig:grpdelay}\,(c)) confirm this time–frequency mapping and demonstrate continuous spectral tuning via pump–seed delay.

As illustrated in Fig.~\ref{fig:pulsesim}, the increased dispersion provided by the SF1 glass effectively maps the broadband seed spectrum (approximately 680–1000\,nm) onto the temporal domain. This mapping extends over a time window of $\sim$1500\,fs, enabling selective amplification of narrow spectral regions through controlled pump–seed delay. The resulting improvement in spectral selectivity is consistent with the expected behavior of chirped-pulse optical parametric amplification \cite{Cerullo2003}.

\begin{figure}
    \centering
    \includegraphics[width=0.9\linewidth]{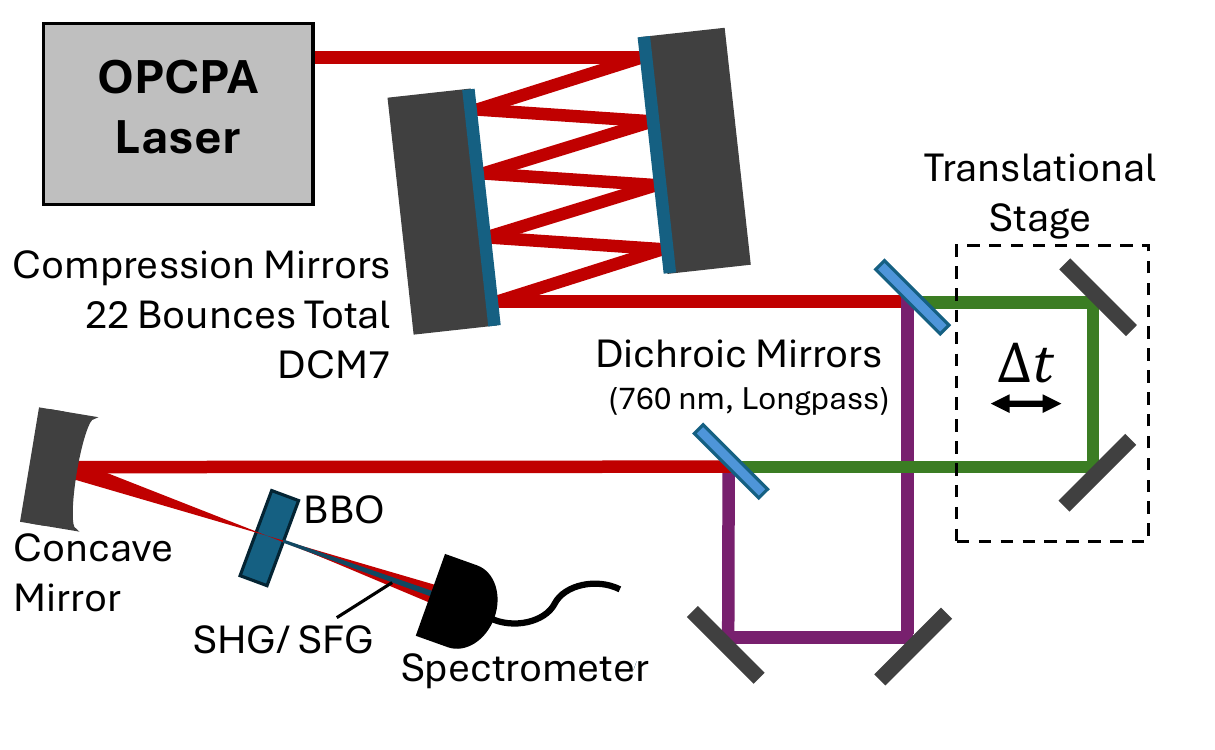}
    \caption{Configuration of the cross-correlator used to characterize the two-color pulses generated in the PBS-based OPCPA scheme.}
    \label{fig:ccdiagram}
\end{figure}

To verify the generation and temporal controllability of the two-color output, cross-correlation measurements were performed using a two-arm interferometric setup \cite{Diels1985} (see Fig.~\ref{fig:ccdiagram}). The two spectral components were separated using a dichroic mirror and recombined after introducing a variable delay. Nonlinear mixing in a BBO crystal produced second-harmonic (SHG) and sum-frequency generation (SFG) signals, which were spectrally resolved.

\begin{figure}
    \centering
    \includegraphics[width=0.9\linewidth]{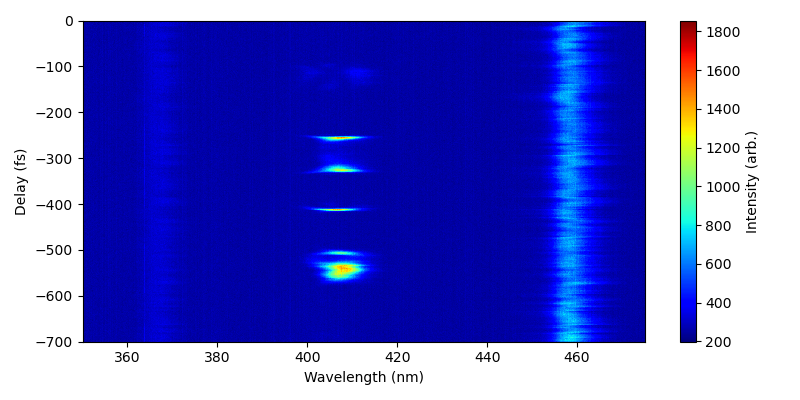}
    \caption{Cross-correlation scan of the two-color OPCPA output using the PBS-based dispersive scheme. Second-harmonic signals from the 730\,nm and 920\,nm components appear at 365\,nm and 460\,nm, respectively, while sum-frequency generation at 410\,nm is observed only under temporal overlap of the two pulses.}
    \label{fig:ccscan}
\end{figure}

Distinct SHG signals are observed at 365\,nm and 460\,nm, corresponding to the 730\,nm and 920\,nm components (Fig.~\ref{fig:ccscan}), respectively, while SFG at 410\,nm appears only under temporal overlap. This confirms the simultaneous generation and independent temporal control of the two spectral components. The cross-correlation trace exhibits a main peak with a duration on the order of $\sim$60\,fs. 

For the measured spectral bandwidths of approximately 20\,nm (FWHM), the Fourier-transform-limited pulse durations are on the order of $\sim$40--60\,fs, depending on the central wavelength of the respective spectral component. The observed cross-correlation width is therefore consistent with near-transform-limited pulses within experimental uncertainty, noting that the measurement reflects the convolution of both pulse durations and may include residual dispersion. In addition to the main feature, a series of satellite peaks with significantly shorter apparent durations is observed.

The origin of this multi-peak structure was investigated further. Notably, the temporal spacing and relative amplitude of the satellite peaks were found to be largely independent of the selected wavelengths and of the number of reflections on the chirped mirrors, suggesting that they do not arise from the intrinsic pulse structure or compression conditions. This observation points to the cross-correlation setup itself as the source of the additional temporal features. In particular, the use of dichroic beam splitters (Thorlabs DMLP760) for spectral separation and recombination introduces dispersive and partially reflective interfaces that may generate weak pulse replicas.

\begin{figure}
    \centering
    \includegraphics[width=0.9\linewidth]{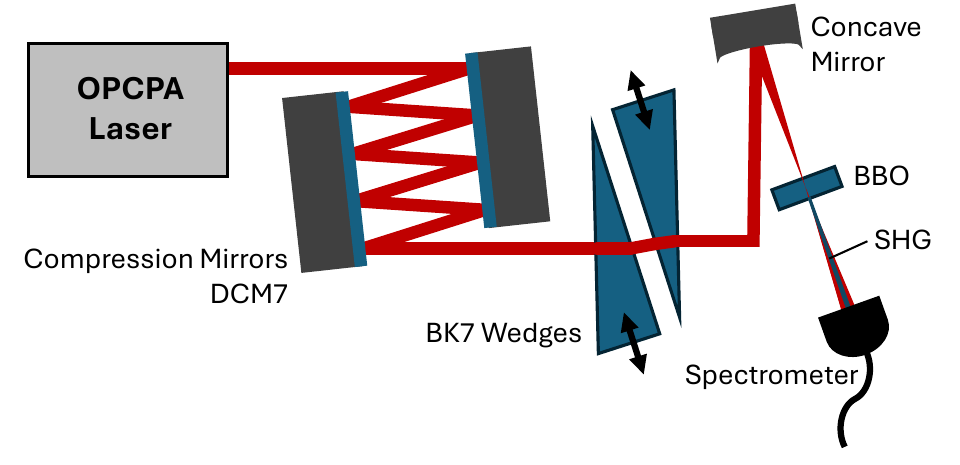}
    \caption{Configuration used to perform D-scan measurements on the laser output. Each wedge was mounted on a translational stage, allowing fine adjustment of the inserted dispersion.}
    \label{fig:dscandiagram}
\end{figure}

\begin{figure}
    \centering
    \includegraphics[width=0.9\linewidth]{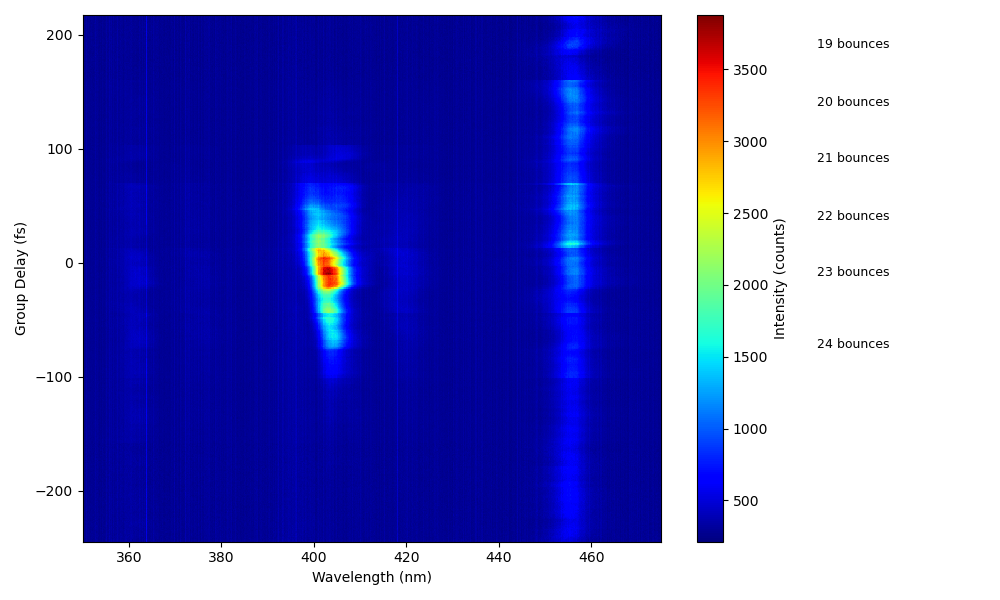}
    \caption{Dispersion scan of the two-color OPCPA output at 730\,nm and 920\,nm using the PBS-based dispersive scheme. Between successive wedge scans, the number of reflections on the chirped mirrors was adjusted to extend the accessible dispersion range. The corresponding number of reflections is indicated on the right.}
    \label{fig:dscan}
\end{figure}

To test this hypothesis, independent pulse characterization was performed using a dispersion-scan (D-scan) technique \cite{Miranda2012}, as shown in Fig.~\ref{fig:dscandiagram}. In this approach, the OPCPA output is analyzed without beam splitting or recombination. The pulses were compressed using chirped mirrors and subsequently propagated through a pair of BK7 wedges to introduce a variable amount of dispersion. The resulting second-harmonic spectra were recorded as a function of inserted glass thickness, thereby scanning the wavelength-dependent group delay and effectively varying the relative timing of different spectral components within the pulse. To extend the accessible dispersion range, the number of reflections on the chirped mirrors was varied between successive wedge scans. The retrieved D-scan traces (Fig.~\ref{fig:dscan}) yield a pulse duration of $\sim 60$\,fs within experimental uncertainty. Notably, no satellite features are observed in the D-scan traces, indicating that the multi-peak structure seen in the cross-correlation does not originate from the OPCPA output pulses, but is introduced by optical elements within the cross-correlator prior to the nonlinear interaction.

A possible source of these artifacts are the dichroic mirrors used for spectral separation and recombination, which may introduce weak secondary reflections and dispersion effects due to their multilayer coating structure. The comparatively narrow width of the individual peaks is attributed to the large frequency separation between the two spectral components, which leads to a rapidly varying temporal modulation in the nonlinear signal. Consequently, the cross-correlation trace reflects both the pulse envelope and interference between the two frequency components. Replacement of these optics with low-dispersion or ultrafast-optimized beam splitters is expected to suppress these artifacts in future implementations.

As a further characterization of the two-color OPCPA output, a cross-correlation measurement was performed using a cold target recoil ion momentum spectrometer (COLTRIMS) \cite{Hubele2015, Fischer2019} in place of the optical detection channel (BBO crystal and spectrometer). In this configuration, ionization of trapped $^6$Li atoms was used as the nonlinear detection process. The atoms were confined in an all-optical trap, as described in Ref.~\cite{Sharma2018}, with a valence electron prepared in either the $2^{2}S_{1/2}$ ground state or the $2^{2}P_{3/2}$ excited state. The OPCPA output was tuned to produce two spectral components centered at 735\,nm (45\,mW) and 840\,nm (65\,mW), and the electron signal was recorded as a function of the relative delay between the two colors.

\begin{figure}[tb]
    \centering
    \includegraphics[width=0.9\linewidth]{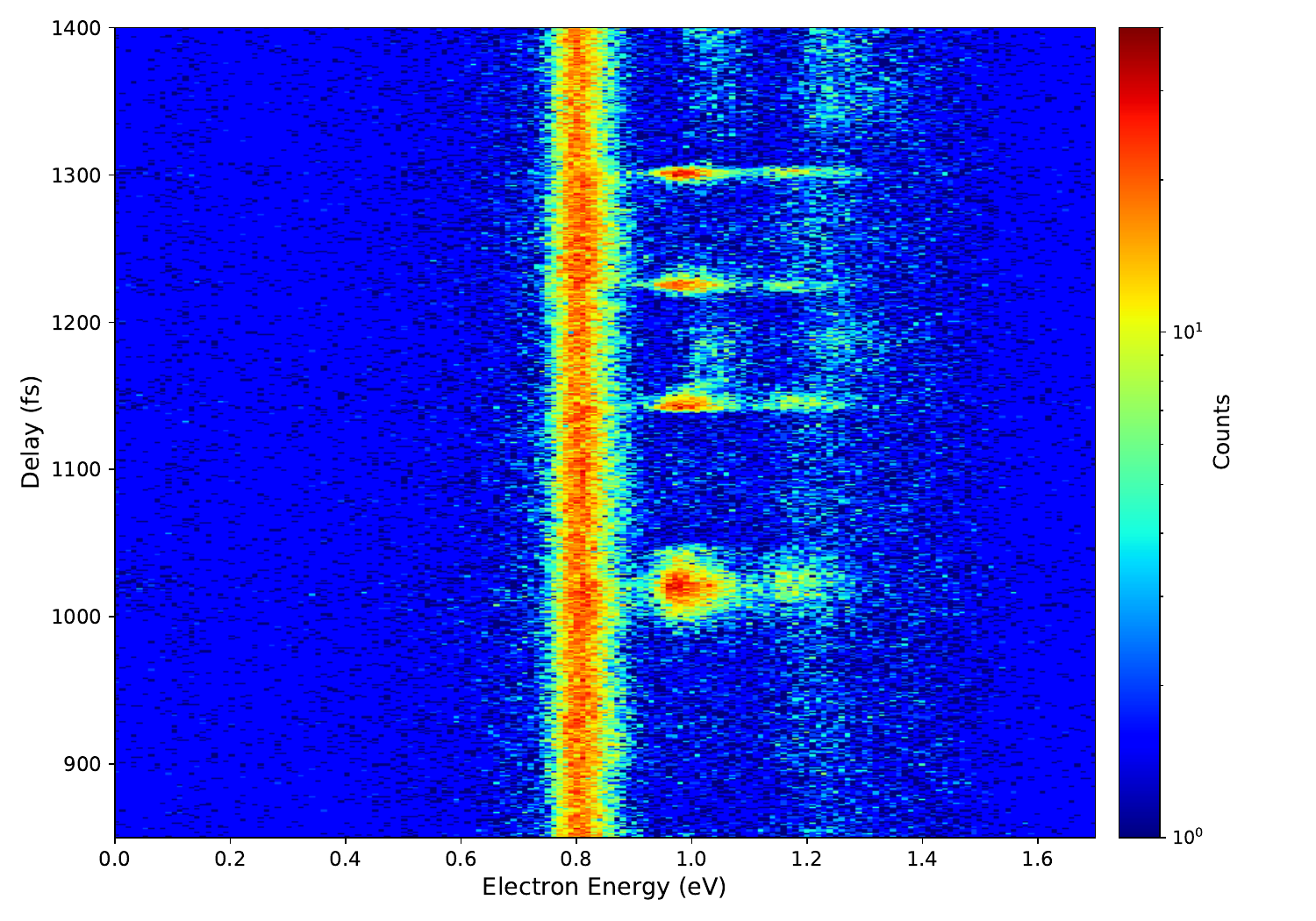}
    \caption{Cross-correlator scan performed using a COLTRIMS setup as the detection spectrometer, with Li atoms as the target. Shown are electron energies resulting from three-photon ionization of the $2^{2}P_{3/2}$ state of Li in an all-optical trap (AOT).}
    \label{fig:nrg}
\end{figure}

The resulting electron energy spectra, shown in Fig.~\ref{fig:nrg}, exhibit a pronounced structure that closely resembles the nonlinear signal observed in the optical cross-correlation measurements (Fig.~\ref{fig:ccscan}). In particular, a dominant feature around 0.8\,eV is observed over the full delay range, while additional peaks appear only when both spectral components temporally overlap. This demonstrates that the atomic ionization process provides a nonlinear response to the two-color field, enabling a direct mapping between optical pulse overlap and electron emission probability.

The observed spectral features can be understood in terms of multiphoton ionization pathways involving different combinations of 735\,nm and 840\,nm photons. The 0.8\,eV feature is consistent with ionization from the $2^{2}P_{3/2}$ state via absorption of three 840\,nm photons. The peak at 0.98\,eV arises from mixed pathways involving two 840\,nm photons and one 735\,nm photon, while the 1.22\,eV feature is consistent with absorption of one 840\,nm photon and two 735\,nm photons. The requirement of temporal overlap between the two colors for the appearance of these features confirms the role of bichromatic excitation in accessing specific multiphoton ionization channels.

Overall, the presented measurements demonstrate controlled generation and characterization of a two-color femtosecond field with independently tunable spectral components and adjustable relative delay. The combination of optical cross-correlation, dispersion-scan characterization, and COLTRIMS-based ionization measurements provides a consistent picture of the pulse structure and its interaction with matter. Importantly, the observed agreement between optical nonlinear signals and strong-field ionization response highlights the robustness of the two-color OPCPA scheme for bichromatic excitation experiments. These results establish the basis for its application in time-resolved studies of atomic and molecular dynamics in the multiphoton regime.

\section{Conclusion}

We have demonstrated a chirped-seed noncollinear optical parametric amplification scheme that enables stable generation of independently tunable two-color femtosecond waveforms with controllable relative delay. The approach combines broadband parametric amplification with controlled time–frequency mapping in a dispersively stretched seed pulse, providing a flexible route to structured ultrafast fields without requiring active pulse shaping in the Fourier domain. 

The resulting architecture is compatible with high-repetition-rate operation and supports direct integration with nonlinear and strong-field detection schemes, offering a compact and adaptable source for multidimensional light–matter interaction studies. In particular, the ability to engineer temporally separated spectral components within a single amplification platform opens new possibilities for tailored excitation in atomic, molecular, and correlated electron dynamics. Future developments will focus on improving temporal fidelity of multi-arm detection schemes and extending the approach toward higher pulse energies and broader spectral coverage.

\section*{Acknowledgements}

This work was supported by the U.S. National Science Foundation under Grant No. PHY-2207854.

%

\end{document}